# Characterization of Exogenic Boulders on Near-Earth Asteroid (101955) Bennu from OSIRIS-REx Color Images

Lucille Le Corre[1][2], Vishnu Reddy[3], William F. Bottke[4], Daniella N. DellaGiustina[3], Keara Burke[3], Jennifer Nolau[5], Robin B. Van Auken[3], Dathon R. Golish[3], Juan Sanchez[1], Jian-Yang Li[1], Christian d'Aubigny[3], Bashar Rizk[3], Dante Lauretta[3]

Abstract

A small number of anomalously bright boulders on the near-Earth, rubble-pile asteroid (101955) Bennu were recently identified as eucritic material originating from asteroid (4) Vesta. Building on this discovery, we explored the global presence of exogenic boulders on Bennu. Our analysis focused on boulders larger than 1 m that show the characteristic 1-$\mu$m pyroxene absorption band in the four-color MapCam data from the OSIRIS-REx mission. We confirm the presence of exogenic boulders similar to eucrites and find that mixtures of eucrites with carbonaceous material is also a possible composition for some boulders. Some of the exogenic boulders have spectral properties similar to those of ordinary chondrite (OC) meteorites, although the laboratory spectra of these meteorites have a higher albedo than those measured on Bennu, which could be explained by either a grain size effect, the presence of impact melt, or optical mixing with carbonaceous material owing to dust coating. Our Monte Carlo simulations predict that the median amount of OC mass added to the parent body of Bennu is 0.055% and 0.037% of the volume of a 100- and 200-km-diameter parent body, respectively. If Bennu was a uniformly mixed byproduct of parent body and S-type projectiles, the equivalent mass of OC material would be a sphere with a diameter of 36 to 40 m (or a volume of 24,200 to 33,600 m$^3$). The total amount of OC material in the interior of Bennu estimated from the MapCam data is slightly higher (91,000–150,000 m$^3$).

[1] Corresponding author lecorre@psi.edu
[2] Planetary Science Institute, 1700 East Fort Lowell Road, Tucson, AZ 85719, USA
[3] Lunar and Planetary Laboratory, University of Arizona, 1629 E University Blvd, Tucson AZ 85721-0092, USA
[4] Department of Space Studies, Southwest Research Institute, 1050 Walnut St, Suite 400, Boulder, CO 80302, USA
[5] University of Central Florida, Department of Physics, 4000 Central Florida Blvd. Orlando, FL 32816

# 1. INTRODUCTION

Catastrophic collisions between planetary bodies are among the most fundamental geological processes in our solar system. Collisions are thought to have created planetary and asteroidal moons and caused mass extinctions on Earth. The visible evidence for collisions is the large and small impact craters that dot planetary surfaces. Impactors have delivered exogenic materials that have enriched native regolith with minerals, volatiles, and organic material.

Our understanding of this medley of solar system materials comes from five main sources: (1) laboratory studies of brecciated meteorites and Apollo lunar samples that contain exogenic xenoliths (Baedecker et al. 1973; Brilliant et al. 1992); (2) ground-based, rotationally resolved spectral observations of main belt and near-Earth asteroids that show evidence of surface variegation (e.g. Gaffey 1997; Sanchez et al. 2017); (3) ground-based, spatially resolved images from large telescopes revealed albedo variegation on the surface of (2) Pallas (Marsset et al., 2020), (16) Psyche (Viikinkoski et al. 2018; Ferrais et al. 2020), and (10) Hygiea (Vernazza et al. 2020); (**4**) proximal spacecraft observations of asteroids (e.g., Reddy et al. 2012; DellaGiustina et al. 2020a); and (**5**) iron meteorites on the surface of Mars discovered by rovers (Schröder et al. 2008).

The surface of asteroid (4) Vesta, a 525-km differentiated basaltic asteroid, is covered with material from primitive carbonaceous impactors on a hemispherical scale, with a global average of 8 wt % of impactor materials. The howardite, eucrite and diogenite (HED) meteorites from Vesta show evidence of carbonaceous xenoliths that in some cases (e.g., howardite PRA04401) make up 60% of the host meteorite (Herrin et al. 2011). Shepard et al. (2015) proposed the presence of exogenic impactor material on large metallic (M-type) asteroids that have a prominent 3-$\mu$m absorption band indicative of hydration. These "wet" M-type asteroids are thought to contain exogenic carbonaceous chondrites rich in volatiles (OH/$H_2O$) on a hemispherical scale. Recent observations of 252-km M-type asteroid (16) Psyche confirmed this prediction and showed that three-quarters of Psyche's surface is covered by low-albedo carbonaceous material (Shepard et al. 2017; Sanchez et al. 2017; Takir et al. 2017).

Smaller asteroids also appear susceptible to exogenic contamination. The Hayabusa mission to near-Earth asteroid (25143) Itokawa observed a possibly exogenic dark boulder on the surface of this stony (S-type) asteroid (Hirata & Isiguro 2011). In addition, the impact of 2008 TC3 in 2008 delivered the Almahata Sitta meteorites, which showed that this small 4.1 m asteroid was made of 70—80% ureilite and 20—30% enstatite and ordinary (H and L) chondrites (Bischoff et al. 2010). And finally, the telescopic optical navigation camera onboard the Hayabusa2 spacecraft made observations of distinct bright boulders on the surface of carbonaceous asteroid (162173) Ryugu, and some of them have near-infrared color spectra consistent with ordinary chondrites meteorites (Tatsumi et al., 2020).

All these lines of evidence point to a size-independent contamination of asteroid surfaces ranging from 500 km down to 4 m in size. In addition, analyses of Apollo samples show that on average, the lunar regolith contains 3.5 wt.% exogenic material (Baedecker et al. 1973) and 1–2 wt.% carbonaceous chondrite material (Brilliant et al. 1992).

The carbonaceous near-Earth asteroid (101955) Bennu, a rubble pile made from the reaccumulated fragments of a larger parent body, is the target of NASA's OSIRIS-REx (Origins, Spectral Interpretation, Resource Identification, and Security–Regolith Explorer) mission, which is bringing back a sample of regolith to Earth (Lauretta et al. 2017, 2021). Using multispectral data collected by the MapCam imager of the OSIRIS-REx Camera Suite (OCAMS) (Rizk et al. 2018), DellaGiustina et al. (2020a) discovered six exogenic basaltic boulders on the surface of Bennu thanks to their distinctive spectral shape and higher albedo relative to Bennu's overall dark regolith (e.g., DellaGiustina et al. 2019). These exogenic boulders have the characteristic absorption band at ~1 $\mu$m due to the mineral pyroxene, which is detected in three of the four MapCam color filters. This absorption feature has been independently confirmed by observations made using the OSIRIS-REx Visible and InfraRed Spectrometer (OVIRS) (Reuter et al. 2018) that showed the complete 1-$\mu$m band, as well as the associated 2-$\mu$m absorption band expected for pyroxenes. Based on the analysis of OCAMS and OVIRS data, DellaGiustina et al. (2020a) concluded that the exogenic basalts, in the form of HED meteorites (in particular eucrites), were incorporated into the parent body of Bennu following their liberation from Vesta, the largest source of basalts in the asteroid belt.

In this paper, we focus on the analysis of the four-color MapCam data of a broader set of potential exogenic boulders to constrain their color properties and identify their meteorite affinities. Although our color dataset lacks the spectral resolution of the OVIRS instrument, the increased spatial resolution of our data helps avoid spectral 'contamination' from Bennu's background terrain. We analyzed color ratios and spectrally matched them with resampled laboratory spectra of meteorites to draw our conclusions.

## 2. DATA PROCESSING

We used a multi-band mosaic built with images at 25 cm/pixel from the OCAMS MapCam camera (Figure 1A&B) to analyze the color properties of boulders (DellaGiustina et al. 2020a). We also used a mosaic constructed from OCAMS PolyCam panchromatic images at ~5 cm/pixel (Bennett et al., 2020) to study the morphology of the boulders with higher-resolution images (Figure 2A, 2B and 2C). The absolute uncertainty for the OCAMS calibrated images is 5% and the relative uncertainty is 1% (Golish et al. 2020a). For the MapCam image mosaic, camera pointing was adjusted so that the image-to-image registration was improved to facilitate extraction of color spectra. All images were also registered to a shape model of Bennu (Barnouin et al. 2020) for the creation of controlled mosaics allowing comparison between MapCam color images and PolyCam panchromatic images. The registration method is described in detail in Bennett et al. (2020) and in the Supplementary Materials from DellaGiustina et al. (2020b). Photometric correction was applied to the radiance factor ($I/F$) color data after calibration (Golish et al. 2020a) to convert to a standard viewing geometry using the photometric models from Golish et al. (2020b).

Both mosaics were assembled with an equirectangular projection using a Bennu radius of 250 m and based on a high-resolution shape model of Bennu (version v28 of the shape model from Barnouin et al. (2020)) with a ground sample distance (GSD) of 80 cm. Global mosaics were created using applications from the ISIS3 software as described in DellaGiustina et al. (2020a, 2020b) for MapCam data and Bennett et al. (2020) for PolyCam data. The color mosaic consists of four bands in the visible and near-infrared with filters b' (473 nm), v (550 nm), w (698 nm) and x (847 nm). Boulders with a distinct shape were mapped as polygons using the MapCam image mosaic in a Geographic Information System, as described in the Supplementary Materials from DellaGiustina et al. (2020b). The median value of I/F was extracted in each of the color filters for all the 1593 mapped boulders studied in this work.

## 3. BOULDER CLASSIFICATION

The first criterion for classification of the boulders in our dataset was the presence of an absorption feature centered around 1 µm attributed to pyroxene and detected as a reflectance peak in the w filter (698 nm), $W_{peak}$. We call the group of boulders that bear this feature PYR. If instead there is an absorption band at this wavelength (drop in reflectance in the w filter), as observed in the spectra of carbonaceous chondrites, it could be due to the presence of iron oxides in clay minerals that were formed by aqueous alteration of the parent asteroidal surface. We call the corresponding group of boulders PHY due to the possible presence of phyllosilicates. Analysis of OVIRS data by DellaGiustina et al. (2020a) confirmed the presence of the 1-$\mu$m band due to pyroxene for six of the boulders or clasts included in our PYR group. A formula based on the work presented in Bell and Crisp (1993) and Bell et al. (2000) was employed to compute the band depth of spectral feature at 698 nm used for distinguishing between the PYR ($W_{peak} > 0$) and PHY ($W_{peak} < 0$) groups. The formula is as follows:

$$W_{peak} = (R_w/(0.50168*R_v+0.49831*R_x))-1 \quad (1)$$

with $R_w$, $R_v$, and $R_x$ corresponding to the reflectance values for the filter w, v, and x, respectively.

Within these two groups, boulders were further divided using the b'/v ratio values to separate boulders with b' reflectance lower than v reflectance (PYR1, PHY1), which is typical of pyroxene-rich material, and those with b' reflectance higher than v reflectance (PHY2, PYR2). These two last groups exhibit color spectra with an overall blue (negative) spectral slope, more consistent with the moderately blue average spectrum of the global surface of Bennu (DellaGiustina et al. 2020b).

These groups were additionally subdivided using the v reflectance level with a threshold value of 0.049. This value corresponds to the median value of the bimodal distribution observed for all mapped boulders (DellaGiustina et al., 2020b). For boulders with reflectance in the v filter lower than 0.049, we defined the subgroups PYR1a (11 boulders), PYR2a (424 boulders), PHY1a (1 boulder), and PHY2a (542 boulders). For the boulders with v reflectance higher than 0.049, we defined the subgroups PYR1b (170 boulders), PYR2b (275 boulders), PHY1b (46 boulders), and PHY2b (124 boulders). A summary for the different boulder groups and subgroups and their characteristics are presented in Table 1.

*Table 1.* Summary of the boulder classification with the resulting groups and sub-groups.

| Group | Sub-group | 1st criterion | 2nd criterion | 3rd criterion | Number of boulders |
|---|---|---|---|---|---|
| PYR1 | PYR1a | $W_{peak} > 0$ | b' < v | v < 0.049 | 11 |
|  | PYR1b |  |  | v > 0.049 | 170 |
| PYR2 | PYR2a |  | b' > v | v < 0.049 | 424 |
|  | PYR2b |  |  | v > 0.049 | 275 |
| PHY1 | PHY1a | $W_{peak} < 0$ | b' < v | v < 0.049 | 1 |
|  | PHY1b |  |  | v > 0.049 | 46 |
| PHY2 | PHY2a |  | b' > v | v < 0.049 | 542 |

| | PHY2b | | | v > 0.049 | 124 |

The subgroup PYR1b included the greatest number of boulders with a $W_{peak}$ detected. Within this subgroup, there is a distinct separation between a group of six boulders with a clear pyroxene signature having $W_{peak}$ > 10% and 164 boulders with a much weaker feature having $W_{peak}$ < 4%. Boulders with similar spectral shape as the PYR1b group but with lower reflectance, classified as PYR1a, are much scarcer and exhibit a $W_{peak}$ only up to 1.4%. In contrast, PYR1b boulders have a $W_{peak}$ that can reach 41% for the brightest (at 550 nm) boulder of the group. All the mapped boulders and their corresponding classification are presented in Table 2.

Tatsumi et al. (submitted) also identified exogenic material on Bennu's surface, using a different method based on the detection of the 1-$\mu$m absorption band computed from the x-filter data. They searched for exogenic material using binned images and set a minimum threshold of 15% for the band depth value. Consequently, in their work, the objects detected are generally much smaller (~0.1 to 3.5 m) and have overall a stronger olivine-pyroxene signature at 1 $\mu$m than the boulders we classify as PYR1a and PYR1b. There are 13 boulders with a diameter around 1 m or larger from the Tatsumi et al. (submitted) dataset included in our work given that they could be identified and mapped as a distinct rock in the MapCam image mosaic. These boulders, indicated with their ID number from Tatsumi et al. (submitted) in Table 2, correspond to boulders from our subgroups PYR1b (11 boulders) and PHY1b (2 boulders). Tatsumi et al. (submitted) identified these boulders as exogenic candidates based on the detection of an absorption in the x filter and not based on $W_{peak}$, therefore, some of them were not classified as exogenic (i.e., PHY1b) in our work.

*Table 2.* List of the mapped boulders on Bennu indicating the boulder ID #, the central latitude and longitude in degrees, $W_{peak}$ value (band depth of the 1-$\mu$m absorption feature of pyroxene), and assigned classification. Note: Boulders identified in common with the study of Tatsumi et al. (submitted) are indicated with their ID numbers from that paper in parentheses (EX##) next to our boulder ID # in the first column. For the complete dataset of boulders, see attached machine readable table file tab2.txt.

| Boulder ID # | Center latitude (°) | Center longitude (°) | $W_{peak}$ | Classification |
|---|---|---|---|---|
| 0 | -24.2648 | 25.1461 | -0.0007 | PHY2a |

| 1 | −32.5727 | 64.2378 | −0.0047 | PHY2a |
| 2 | −30.5673 | 44.2038 | −0.0016 | PHY2a |
| 3 | −26.8431 | 3.7560 | 0.0025 | PYR2a |
| 4 | −17.7903 | 74.5925 | −0.0022 | PHY2b |
| 5 | −31.4840 | 69.6997 | −0.0050 | PHY2b |
| 6 | −59.1579 | 65.8170 | 0.0079 | PYR2b |
| 7 | −60.4184 | 66.4320 | 0.0030 | PYR2b |
| 8 | −59.6736 | 84.2655 | 0.0078 | PYR2b |
| 9 | −57.2098 | 87.1458 | 0.0045 | PYR1b $W_{peak} < 4\%$ |
| 10 | −48.5582 | 71.8270 | −0.0037 | PHY1b |
| 11 (EX16) | −46.6101 | 68.2008 | 0.1870 | PYR1b $W_{peak} > 10\%$ |
| 12 (EX15) | −50.7354 | 61.3451 | 0.1499 | PYR1b $W_{peak} > 10\%$ |

For each of the subgroups, we computed the average spectrum and compared it to the global average spectrum of Bennu (Figure 3). The average spectrum of each subgroup exhibits higher reflectance values than the average spectrum of Bennu, except for the b' value from the PYR1a subgroup, which is similar to the average value for Bennu (Figure 3A). PYR1a and PYR1b average spectra have a distinct spectral shape compared to Bennu's average spectrum (Figure 3B), with a b' value lower than v, in addition to the presence of the 1-μm absorption band visible as a peak in the w filter ($W_{peak} > 0$), so these boulders are the best candidates for an exogenic origin. Boulders from the PYR2a and PYR2b groups (Figure 3C), with reflectance at b' higher than v, appear more typical of Bennu's background terrain, but the presence of a peak at 698 nm (w filter) up to 2% for PYR2a and 3% for PYR2b could point to regolith mixed with some exogenic contamination or exogenic boulders coated by carbonaceous dust. The possibility of a thin layer of dust on Bennu's boulders is consistent with the findings of Rozitis et al. (2020) and Hamilton et al. (in revision). The absorption feature detected in the PHY subgroups is generally weaker than the reflectance peak at w observed for the PYR groups. The PHY1 boulders (b'<v) include seven boulders with 1 to 15.5% band depth, and boulders in PHY2 class (b'>v) include 28 boulders with 1 to 2% band depth; the remaining boulders in both classes have a band depth < 1%.

## 4. SPECTRAL LIBRARY OF METEORITES

We compared the boulders from the subgroups PYR1a and PYR1b that have a color spectrum indicative of pyroxene to spectra of ordinary chondrites (OCs;

originating from S-type bodies) and HED meteorites obtained in the laboratory and resampled to MapCam filters. Most of the meteorite data that we included in our spectral library for this study were extracted from the RELAB database, but some are laboratory spectra of more recent falls or finds (Reddy et al. 2014; Le Corre et al., 2015). The analyzed spectra were collected from powdered samples with varying grain size (such as < 45 µm, 45–90 µm, 90–250 µm, 250–500 µm) as well as from slabs or surfaces of meteorites. Among the OCs, our library has 21 H chondrites, 20 L chondrites, and 16 LL chondrites, including some spectra of recent falls and some spectra taken on the unaltered or melted/shocked part of the meteorites. Among the HED meteorites, we used 64 spectra of howardites, 139 spectra of eucrites (including some cumulate eucrites, shocked eucrites and melts), and 52 spectra of diogenites (with one olivine-diogenite). Our set of HED spectra also includes data from recent falls in addition to some recent finds. Finally, 19 spectra of laboratory and areal mixtures of Millbillillie (eucrite) and Murchison (CM2) first discussed in Le Corre et al. (2011) were added to our spectral library.

We searched for the best spectral matches to the pyroxene-bearing boulder subgroups (PYR1a and PYR1b). First, we selected the meteorite candidates based on visual comparisons in absolute reflectance, with spectra normalized at 550 nm and at 700 nm to evaluate the spectral shape. In our classification schemes, exogenic boulders and clasts presented in DellaGiustina et al. (2020a) are represented as PYR1b (sites 1, 2, 3, 4, 6 with $W_{peak}$ > 10% and site 5 with $W_{peak}$ < 4%). We also evaluated the boulders' affinity with OCs or HEDs based on the proximity of data points in scatterplots of color parameters such as ratios, relative slope, albedo and $W_{peak}$.

## 5. BOULDER DISTRIBUTION AND MORPHOLOGY

We plotted the spatial distribution of the different classes of PYR1 boulders, which are the most likely candidates for an exogenic origin, on the RGB color map of Bennu (Figure 1A). Most of the PYR1 boulders appear to be distributed at the mid-latitudes with a few near the equator. A lot more exogenic boulders were identified in the equatorial region by Tatsumi et al. (submitted), however, these boulders are smaller than a few meters (corresponding to a few pixels in the images) and are generally much smaller than the exogenic boulders mapped in our work. This difference in the spatial distribution of exogenic boulders between Tatsumi et al. (submitted) and our work could be explained by the effects of mass movements on Bennu. Jawin et al. (2020) proposed that mass movements of material occurred from the mid-latitudes to the equatorial ridge, excavating larger boulders in the mid-latitudes and covering them with a 5-m layer at the equator. This redistribution of material would have filled small craters and accumulated more regolith at the equator. Therefore, the paucity of larger exogenic boulders in the equatorial region might be linked to the observations of patterns of mass movements. Some boulders are distributed along circular features that may be eroded crater rims, between 0–50°E and 20–60°N. There is also a group of boulders scattered radially around what appears to be an ancient impact crater centered at 60°W, 45°N. This pattern most likely corresponds to redistribution of boulders after an impact rather than the remaining material from the crater-forming impactor itself. Multi-meter exogenic boulders could not have been delivered directly to Bennu via impact without disrupting it, though it is plausible that Bennu was assembled from larger precursors that did go through disruption. It is more likely these boulders were delivered to Bennu's parent body by an impactor (DellaGiustina et al., 2020a), and are therefore distributed throughout Bennu's interior. The floor of these craters might have been covered by a layer of dark regolith particles from Bennu (via mass wasting or redeposition after the impact) hiding any potential exogenic boulders inside the crater while exposing many others along the crater rim. It is interesting to note that some boulders from the PYR2 classes also show similar distribution patterns (Figure 1B). It could be due to a bias in mapping boulders larger than a few meters that are more readily visible in crater rims than crater interiors where accumulation of regolith can occur due to mass movements. The remaining PYR1 boulders appear to be randomly scattered at the mid-latitudes.

*Figure 1.* Global distribution of the different classes of PYR1 exogenic boulders on Bennu with PYR1a in yellow, PYR1b with $W_{peak}$ > 10% in cyan and PYR1b with $W_{peak}$ < 4% in red (A). The global distribution of boulders for the PYR2a (blue) and PYR2b (cyan) groups (B). For both maps, the background image is a global map of Bennu in natural colors using MapCam data with red as 698 nm, green as 550 nm and blue as 473 nm.

A.

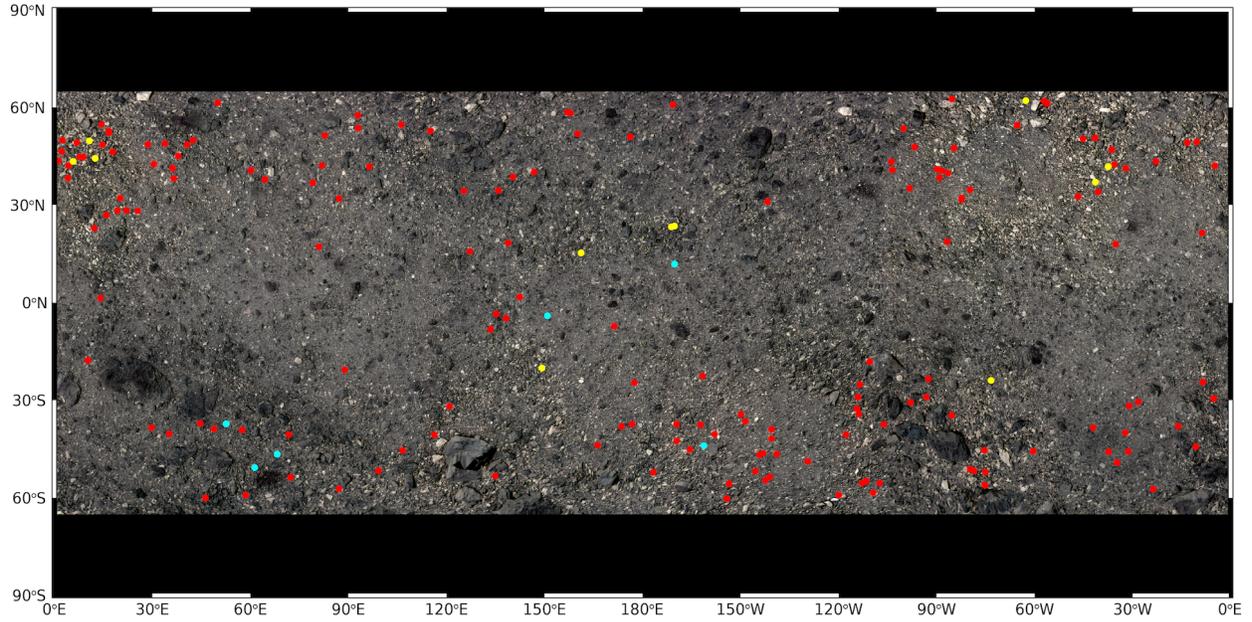

**B.**

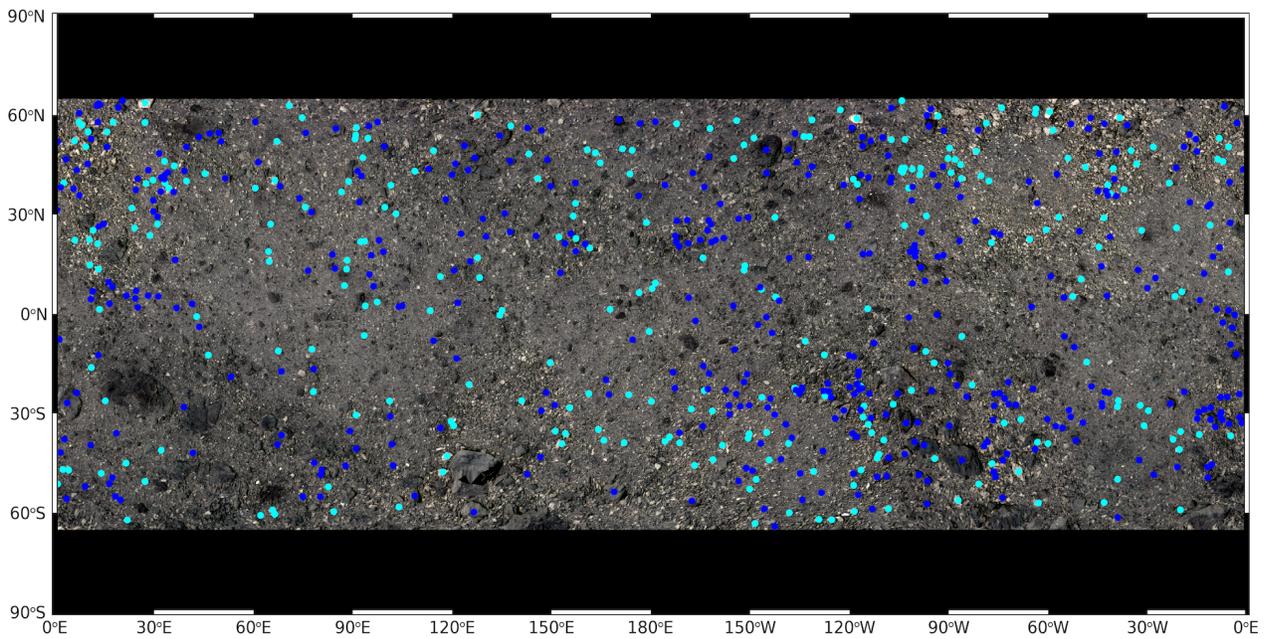

We present close-up images from the PolyCam and MapCam mosaics to show examples of typical boulder morphology and natural colors for a subset of boulders in each PYR1 subgroup (Figure 2A–C). Boulders with albedo and spectral shape most resembling those of pyroxenes (PYR1b) are shown in Figure 2A and 2B. PYR1b boulders are generally much brighter than the background regolith in the PolyCam and MapCam images. These boulders are also redder than their surroundings due to their higher reflectance in the w filter relative to the other filters (which is also true for the PYR1a boulders in Figure 2C). Most of these boulders appear as a single rock standing on Bennu's surface, but some seem partially buried in the regolith, whereas a few others look like clasts embedded within a typical dark boulder found on Bennu (as previously shown in DellaGiustina et al., 2020a,b). An example of this is shown in the largest image of Figure 2B; some clasts are casting a shadow on the rock underneath and could represent debris from a pyroxene-rich rock, whereas others are embedded within the matrix of the dark boulder itself. Alternatively, these clasts could represent material freshly exposed on the surface of an exogenic boulder coated with carbonaceous dust from Bennu after an impact. Some of the PYR1b boulders in Figure 2A and 2B have both bright and greyish surfaces, possibly indicating a partial dust coating. The darker pyroxene-bearing boulders from the PYR1a subgroup are shown in Figure 2C. These boulders appear slightly reddish compared to their surroundings in the RGB images but have also a much lower albedo than the PYR1b boulders; thus, they stand out less from the average dark regolith of Bennu.

*Figure 2.* Close-up panchromatic images from PolyCam (odd column of images) and MapCam RGB color images (even columns) of candidate exogenic boulders on Bennu from the OCAMS instrument. The exogenic candidate boulders are visible at the center in each subimage and most of them are ~2–4 m in their longest axis (all shown at the same scale). PolyCam and MapCam images are aligned for each boulder. For the MapCam images, the color composite is created with red as w (698 nm), green as v (550 nm), and blue as b' (473 nm) filter images. All images are extracted from projected global maps of Bennu and can have some distortions due to projection effects. (A) All the boulders belonging to the PYR1b class that have $W_{peak}$ > 10% with an albedo and spectral shape resembling a pyroxene the most and (B) some of the pyroxene-bearing boulders from PYR1b class with $W_{peak}$ < 4%. The biggest subimage in (B) corresponds to a closeup view of the boulder with bright clasts shown just above this image. This high-resolution image from the mission phase Recon A has a pixel scale of 1.6 cm/pixel. (C) Examples of the much darker boulders from the PYR1a class with $W_{peak}$ < 1.4%.

A.

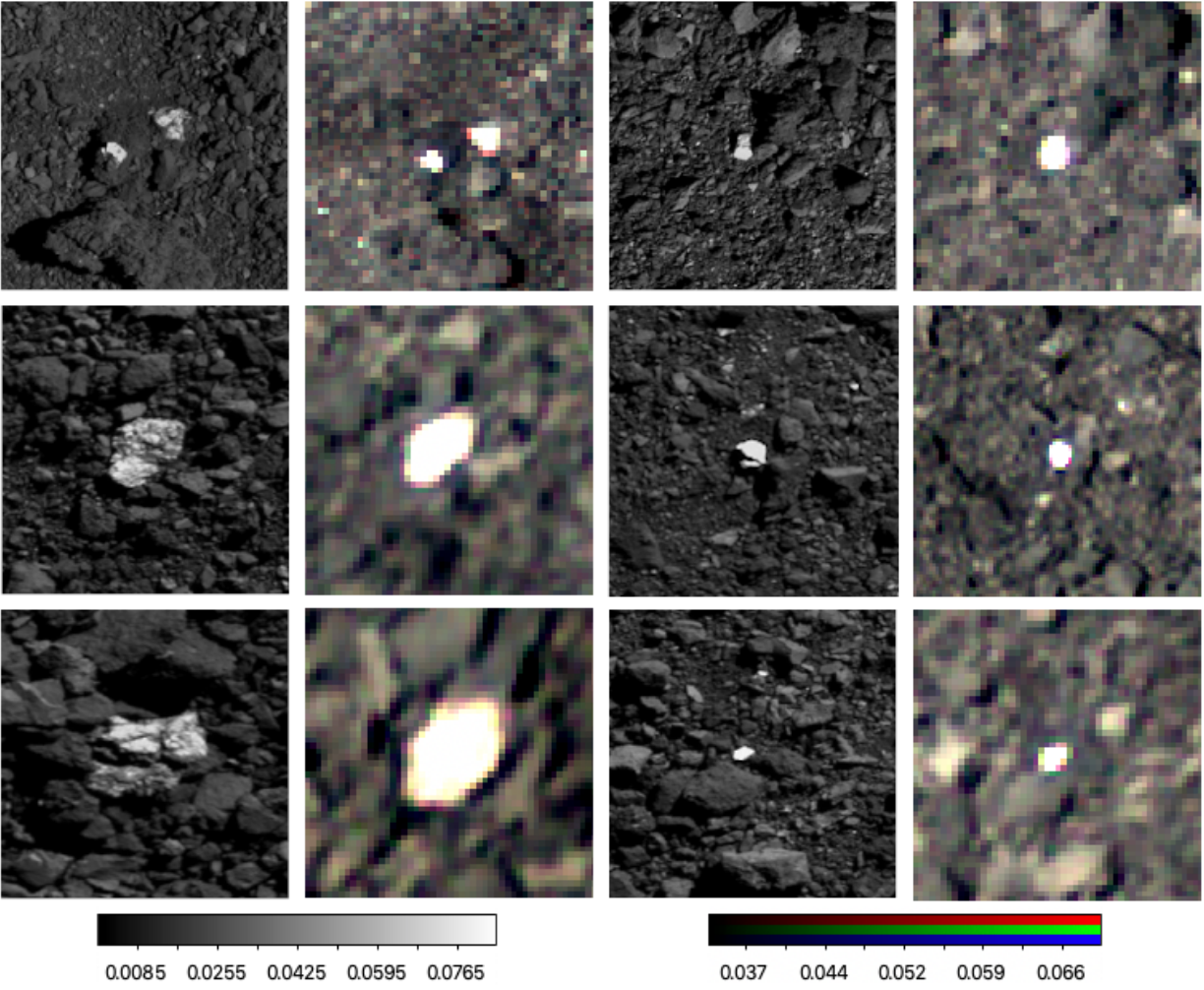

B.

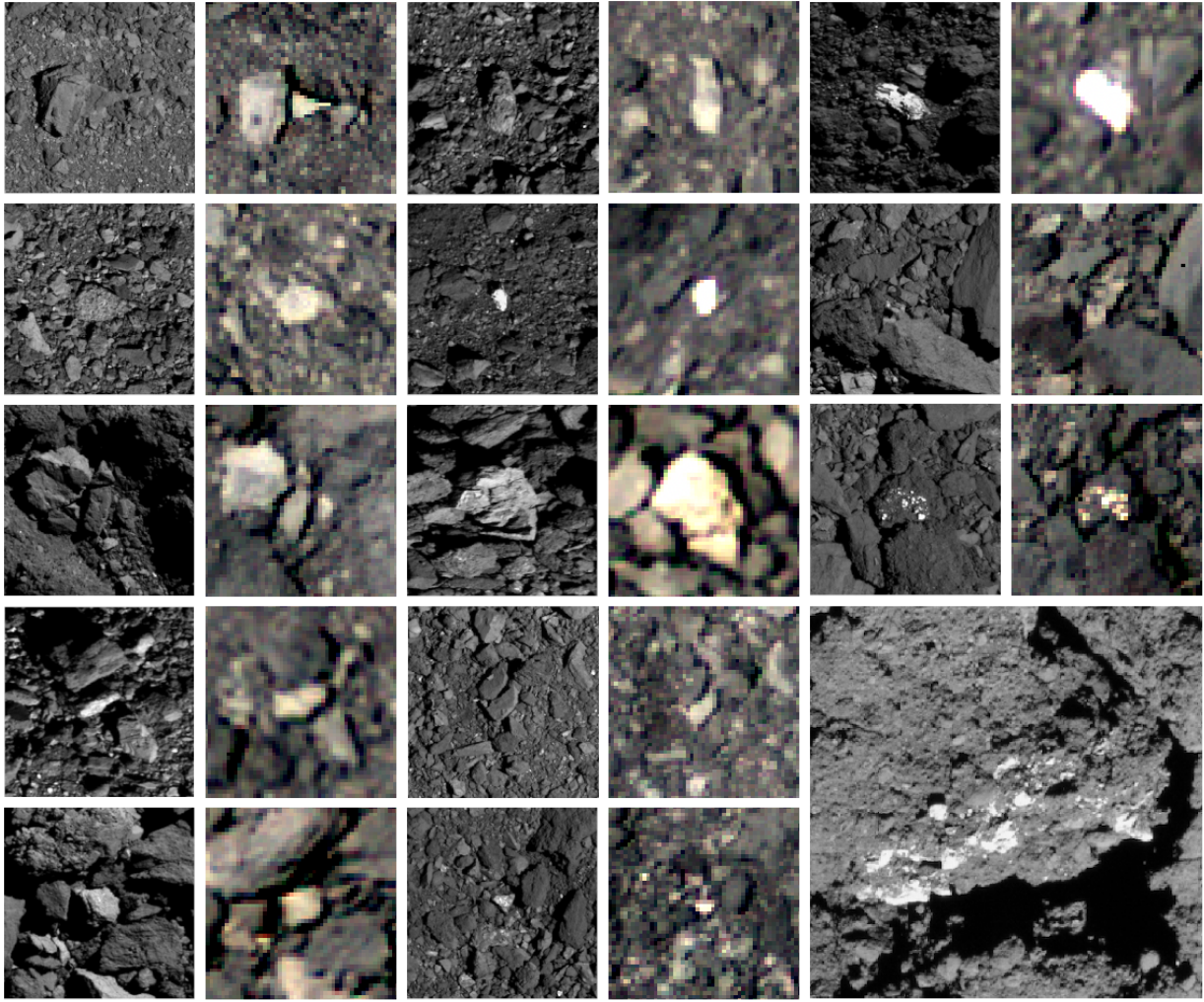

C.

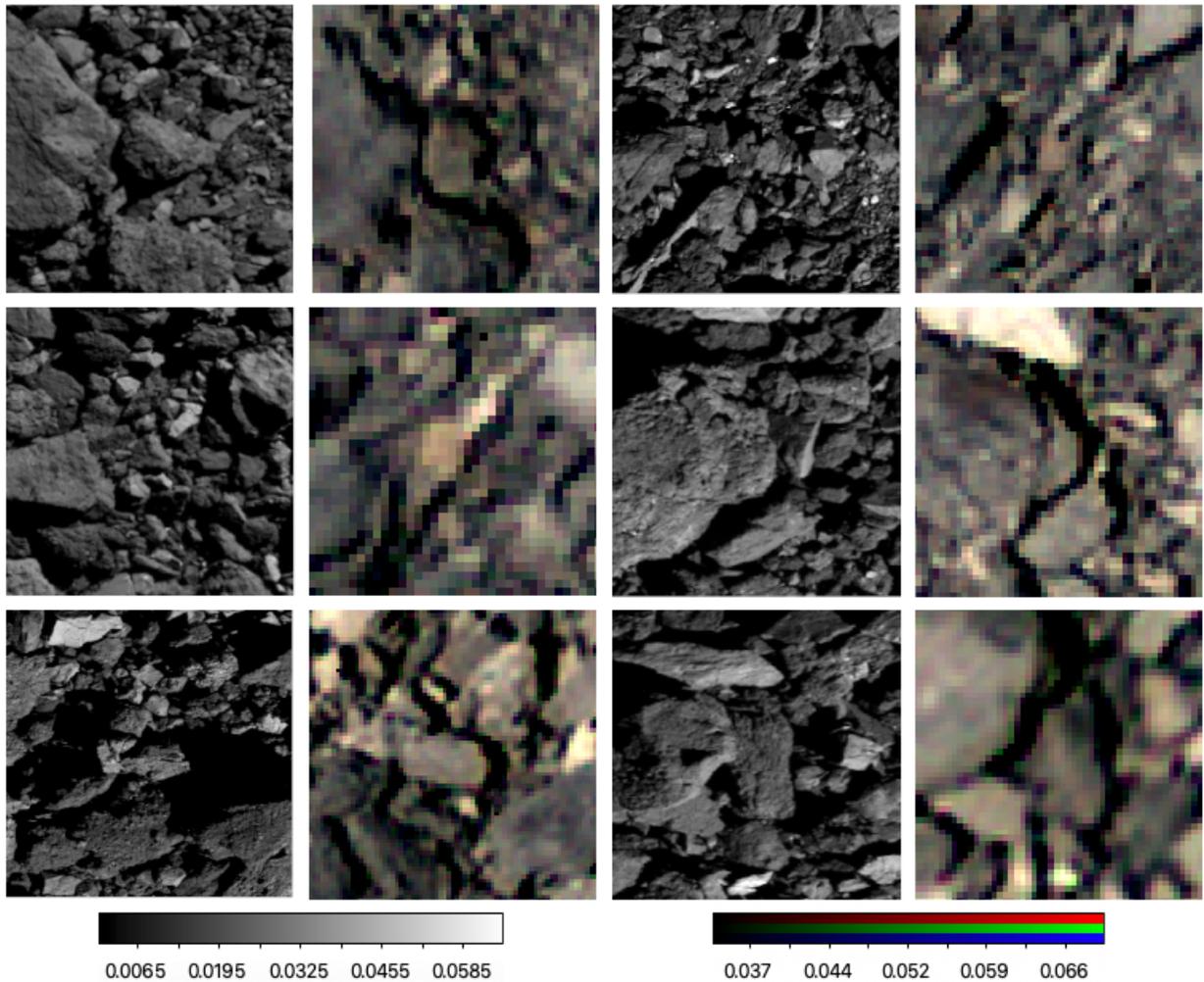

## 6. COMPARISON OF BOULDER COLOR SPECTRA WITH METEORITES

With b' < v, PYR1 boulders show the typical spectral shape of pyroxene-rich material but with a reflectance in the v filter (550 nm) ranging from 0.039 to 0.13, which is significantly lower reflectance than for HEDs (v ~ 0.3) or OCs (v ~ 0.23). This implies the existence of a mechanism that lowers the reflectance without erasing the spectral shape characteristic of pyroxene-rich material. As discussed in section 3, the PYR1 group includes the subgroups PYR1a (with v < 0.049) and PYR1b (with v > 0.049). Figure 3A shows the average four-band color spectrum for PYR1a boulders, PYR1b boulders with $W_{peak}$ > 10%, and PYR1b boulders with $W_{peak}$ < 4%, with error bars indicating the extreme values found for each filter to give a measure of the boulder diversity within each subgroup. Figure 3B has the same spectra normalized at 550 nm. All these groups show the characteristic drop in reflectance in the x filter (centered at 847 nm) due to the mineral pyroxene. They also have a positive b'-v slope, in contrast with the negative slope observed in the global average spectrum of Bennu.

*Figure 3.* (A) Average MapCam spectra of PYR1a and PYR1b (high and low $W_{peak}$ sub-groups separated) with the pyroxene-like absorption feature along with the global average spectrum of Bennu. Error bars indicate the highest and lowest values found in each filter for each group of boulders: dash-dot line for PYR1b with $W_{peak}$ > 10%, grey dashed line for PYR1b with $W_{peak}$ < 4%, and black dashed line for PYR1a. (B) Same data but normalized at 550 nm. (C) Average MapCam spectra of PYR2a and PYR2b normalized at 550 nm compared with the global average spectrum of Bennu.

A.

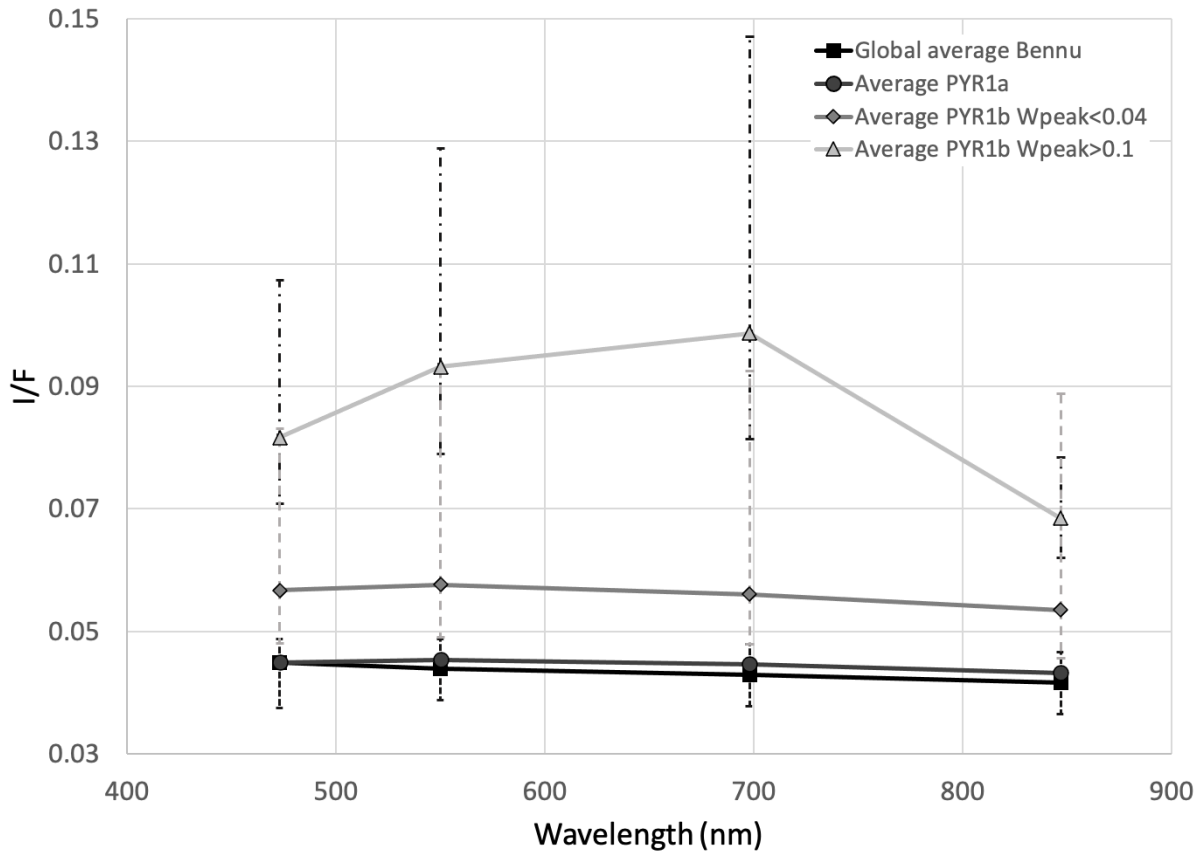

B.

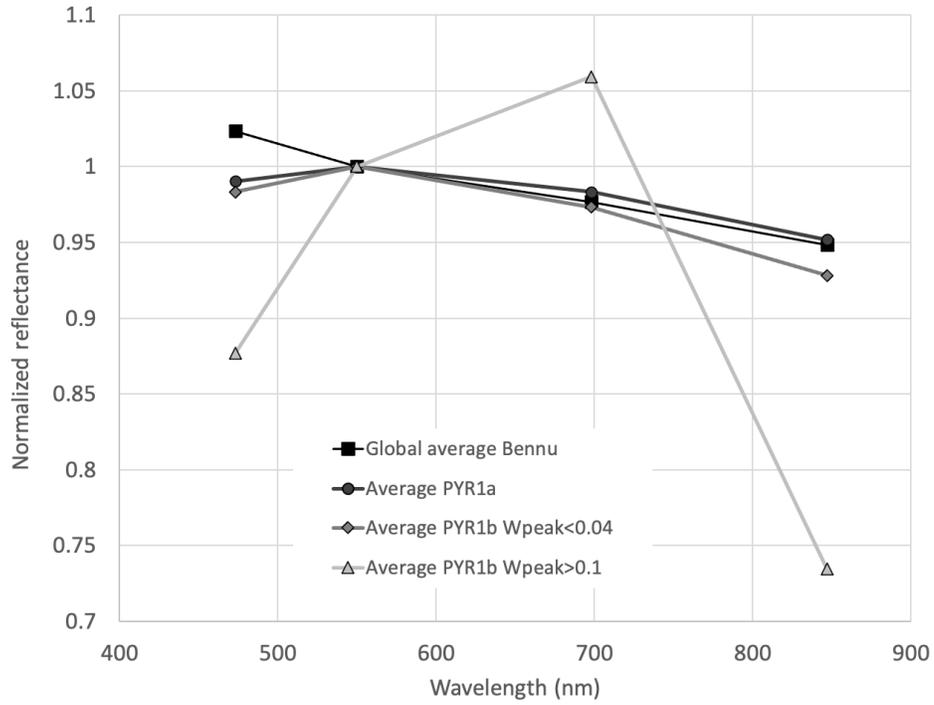

c.

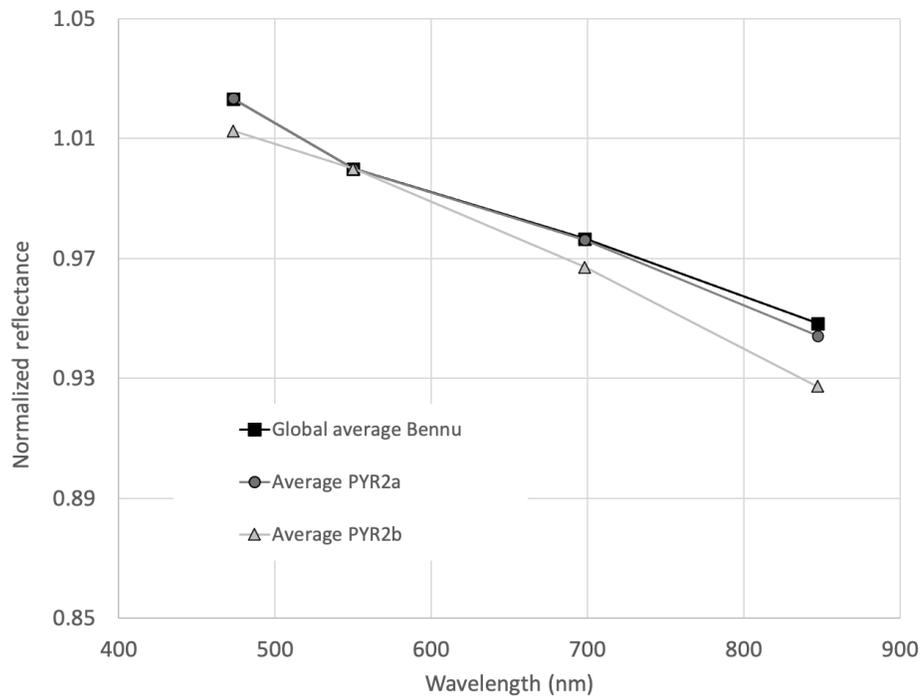

To constrain the meteorite analogs for these candidate exogenic boulders, we surveyed the distribution of meteorite types in the terrestrial collection available via the Meteoritical Bulletin (https://www.lpi.usra.edu/meteor/). Of the 63,900 meteorites in that collection, the two most dominant classes are the OC (85.6%) and HED meteorites (3.6%). Together, these make up 89.2% of all known meteorites, and they both have the characteristic 1-$\mu$m pyroxene absorption band observed in the color spectra of candidate exogenic boulders on Bennu. All other meteorite types that have pyroxene-dominated spectra make up ~1% of the meteorites that fall on Earth. We chose to investigate these two meteorite types (OCs and HEDs) under the assumption that they are the dominant classes among small asteroid populations in near-Earth space and the inner main belt and are likely to have contributed to Bennu's parent body. In addition, some of the bright boulders were identified as HED material by DellaGiustina et al. (2020a) after combining data from MapCam and OVIRS for greater spectral range covering the 2-$\mu$m absorption band diagnostic of pyroxene.

We first investigated the presence of HED meteorites from Vesta on the surface of Bennu to build upon the conclusions of DellaGuistina et al. (2020a). Figures 4A-C show the comparison of the color spectra of PYR1b (with $W_{peak}$ > 10%) boulders with those of selected howardites, eucrites, and diogenites normalized at 550 nm. Only the best fits among the 255 HED spectra based on visual comparisons are shown. Of the three types, eucrites (Figure 4B) seem to match best, in agreement with results from DellaGiustina et al. (2020a). For the v-band albedo, the best match for howardites is in the range 0.14—0.41; for eucrites, 0.08—0.65; and for diogenites, 0.11—0.61. PYR1b boulders with $W_{peak}$ > 10% have a v-albedo range of 0.08-0.13, which is well within the range measured for eucrites. However, we cannot exclude the possibility that exogenic boulders are covered with a thin layer of fine dust from Bennu, which would bring down the albedo (and lower the 1-$\mu$m band depth) enough for some howardites and diogenites to be good matches as well. The only potential meteorite spectral match for the PYR1b with $W_{peak}$ < 4% is a sample of eucrite PCA82501 with a particle size range from submicrons to 1000 $\mu$m, but its 1-$\mu$m band depth is higher that of than these boulders, as is its v albedo (0.23 vs. 0.049—0.092). We did not find a good meteorite match for the PYR1a group as all HEDs have much higher 1-$\mu$m band depths than this group. Further in-depth analysis of the HED mineralogy for a subset of exogenic boulders using the OVIRS data is presented in Tatsumi et al. (submitted).

*Figure 4.* MapCam color spectra of candidate exogenic boulders from the PYR1b class with $W_{peak}$ > 10% and a selection of resampled laboratory spectra of howardite, eucrite, and diogenite (HED) meteorites (A, B, and C) originating

from asteroid (4) Vesta with normalized I/F at 550 nm on the Y-axis and wavelength (nm) on the X-axis.

A.

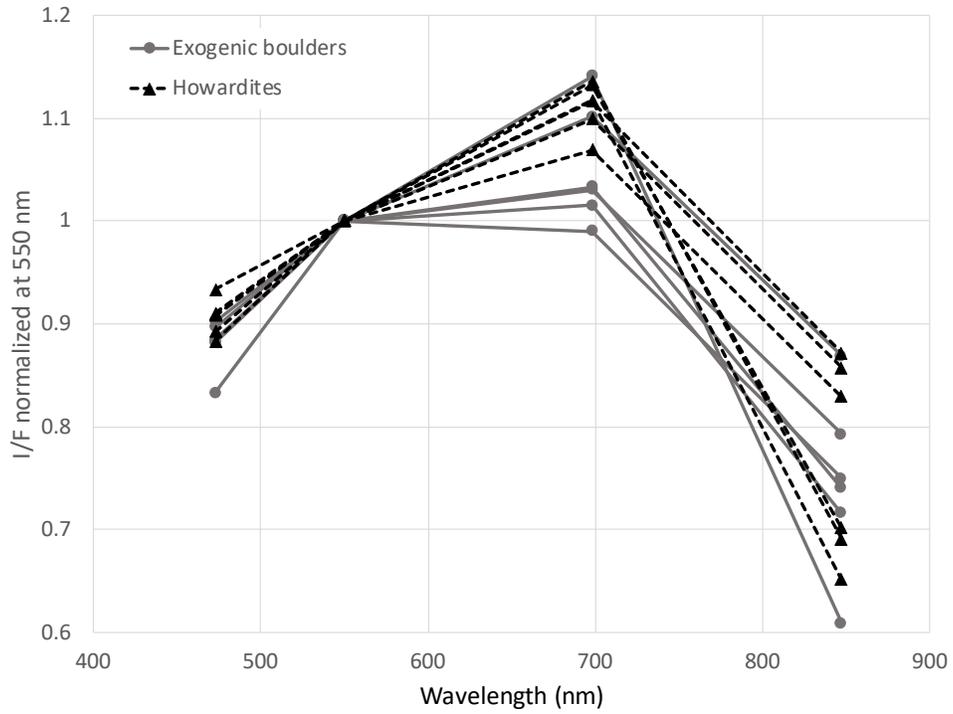

B.

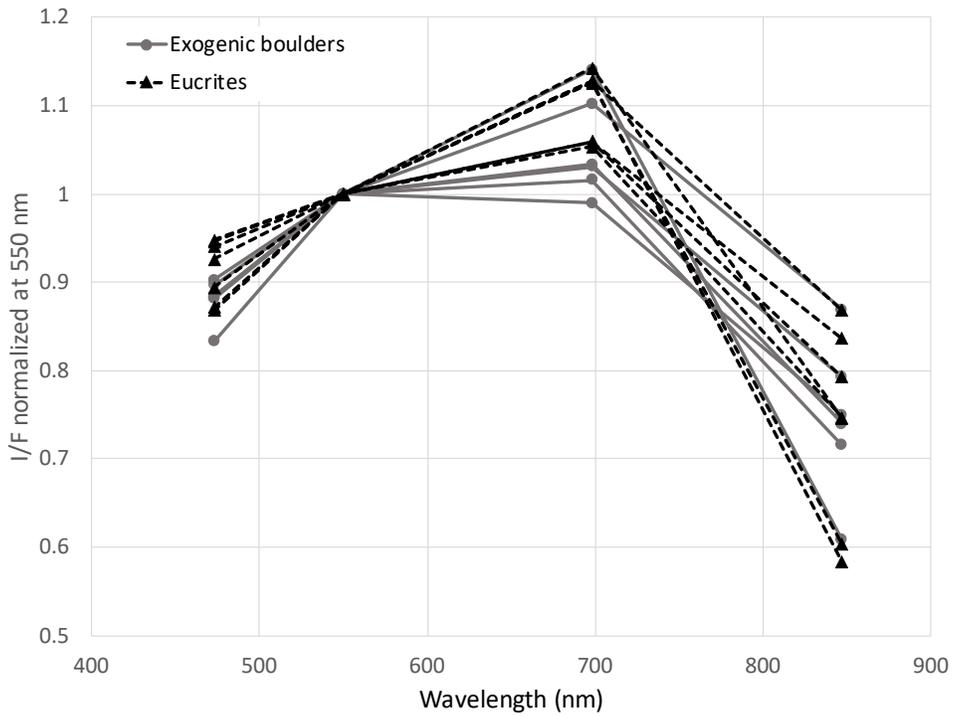

C.

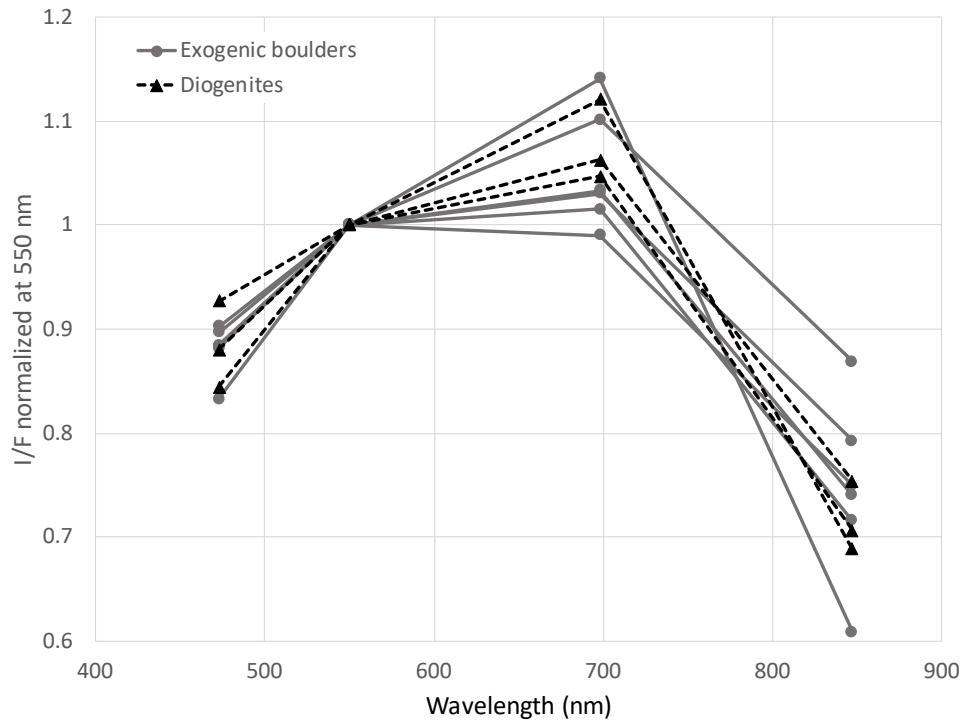

In addition to simple spectral matching, we also investigated color ratio trends to verify the affinity of these candidate exogenic boulders to HEDs. 27% of PYR1a and 44% of the PYR1b boulders are distributed within the range of eucrites in the b'/v vs. v/x plot, and for 5% PYR1b boulders, within the range of howardites as well (Figure 5). The rest of the PYR1a and PYR1b boulders are within the range of values for v/x observed for HEDs but their b'/v ratio is slightly higher than the HEDs' range due to their flatter b'-v slope compared to HED spectra. The higher b'/v ratio could be explained by coating of the boulders with dust from Bennu for example, which would subdue the typical spectral shape for HEDs, or by a different mineralogical composition implying a different spectral shape. The affinity with HEDs is more obvious in Figure 6 (v/x vs. b'/x) where PYR1a and PYR1b boulders plot within the HED trendline and more specifically the eucrites and howardite regions.

*Figure 5.* Scatterplot with the ratio of b'/v filters on the Y-axis and v/x filter on the X-axis for PYR1a and PYR1b boulders. Resampled laboratory spectra of 255 HED meteorites from asteroid (4) Vesta are also shown using different symbols. Color ratios of PYR1a and PYR1b boulders are similar to those of eucrite meteorites, suggesting a compositional affinity.

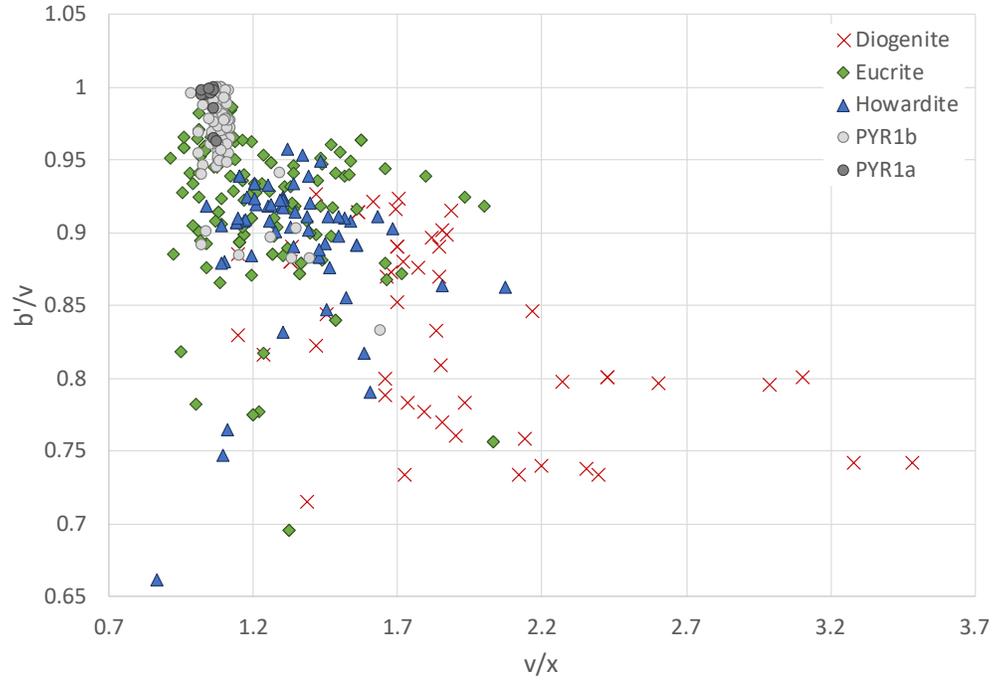

*Figure 6.* Scatterplot with the ratio of v/x filters on the Y-axis and b'/x filter on the X-axis for PYR1a and PYR1b boulders. Resampled laboratory spectra of 255 HED meteorites from asteroid (4) Vesta are also shown using different symbols. Color ratios of PYR1a and PYR1b boulders are similar to those of eucrite meteorites, suggesting a compositional affinity.

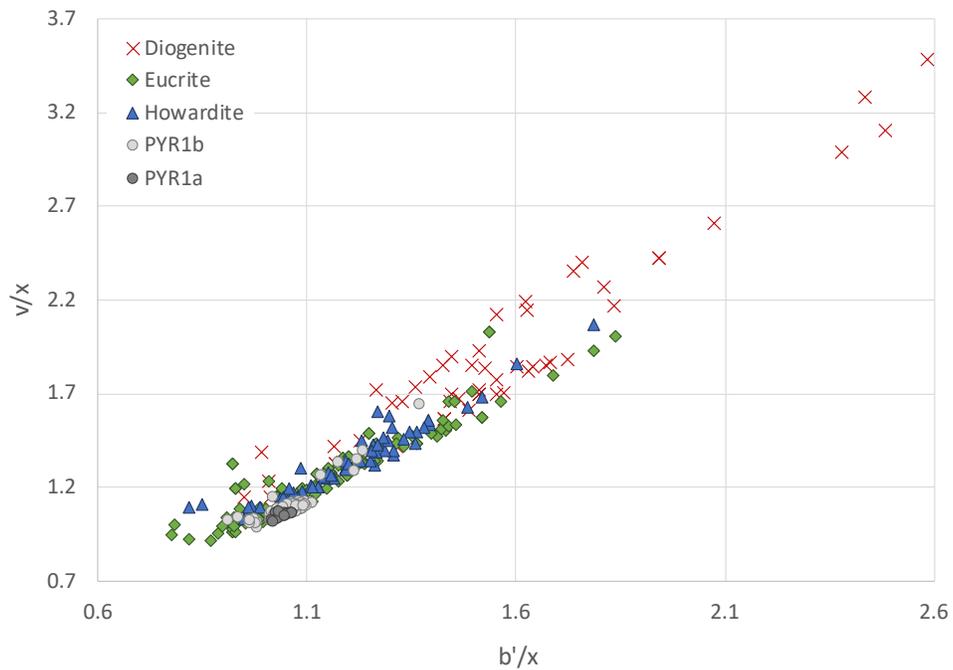

We investigated in more detail the possibility that some boulders could be HEDs covered with a layer of carbonaceous dust from Bennu by comparing them with laboratory spectra of mixtures of eucrite (Millbillillie) and CM2 carbonaceous chondrite (Murchison). Figure 7 shows the color spectra of PYR1b (with $W_{peak} > 10\%$) boulders with the best matches from mixtures of HED and CM2 with different proportions. The v-band albedo of these mixtures (0.08–0.16) and PYR1b boulders (with $W_{peak} > 10\%$; 0.08–0.13) match. The apparent depth of the 1-µm absorption feature is similar to that of some boulders for areal mixtures but too low for the intimate mixtures presented. For lower percentages of CM2 in the intimate mixtures (10–20%), the pyroxene feature better matches the boulders, but the v albedo is too high. None of the mixtures were a good match for boulders from the PYR1a group. The closet match for PYR1b ($W_{peak} < 4\%$) is the intimate mixture with 60% CM2, but higher percentages of CM2 (which are not available) would likely provide a better match with lower v-band albedo and weaker $W_{peak}$. However, the b'/v ratio for the mixture at 60% does not match well.

*Figure 7.* MapCam color spectra of exogenic boulders belonging to the PYR1b class with $W_{peak} > 10\%$, and a selection of resampled laboratory spectra of eucrite (Millbillillie) and CM2 carbonaceous chondrite (Murchison) intimate (30 and 40% CM2) and areal (70 and 80% CM2) mixtures. I/F was normalized at 550 nm. Color spectra of the mixtures match some of the exogenic boulders, suggesting that some might have a dusting of carbonaceous material from Bennu.

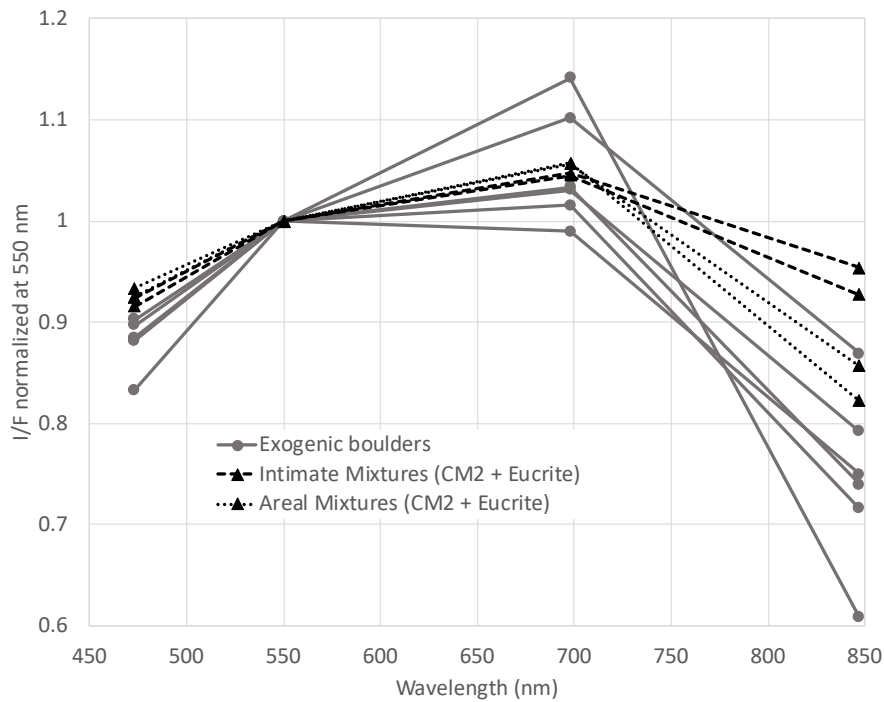

In summary, spectra from brighter boulders show a similar spectral shape than some of the spectra from mixtures, but most low-albedo boulders with the pyroxene absorption band (PYR1a and PYR1b with $W_{peak} < 4\%$) do not match. Intimate mixtures with other type of eucrites might help obtain a better spectral match. This becomes more evident in Figure 8, where we see that the color ratios of HED and carbonaceous chondrite mixtures are offset from most PYR1a and -b boulders, with a slight overlap for some PYR1b boulders. This would imply that most Bennu boulders are pristine eucritic material, and that the subdued absorption bands and lower albedo relative to laboratory spectra might be a particle size effect (regolith vs. slabs). Our interpretation of the OVIRS spectra of exogenic boulders from DellaGiustina et al. (2020a) confirms this: Spectra in Figure 2C of that paper show pyroxene absorption bands with negative spectral slopes. Laboratory spectral measurements of meteorite slabs (bare rock) show negative spectral slope compared to powdered samples (Cloutis et al., 2013).

*Figure 8.* Scatterplot with the ratio of v/x filters on the Y-axis and b'/x filter on the X-axis for PYR1a and PYR1b boulders, and resampled laboratory intimate and areal mixtures of eucrite (Millbillillie) and CM2 carbonaceous chondrite (with 5 and 90% Murchison) shown with different symbols. Color ratios of mixtures are offset from the Bennu exogenic boulders with a slight overlap for some boulders as indicated in the previous plot.

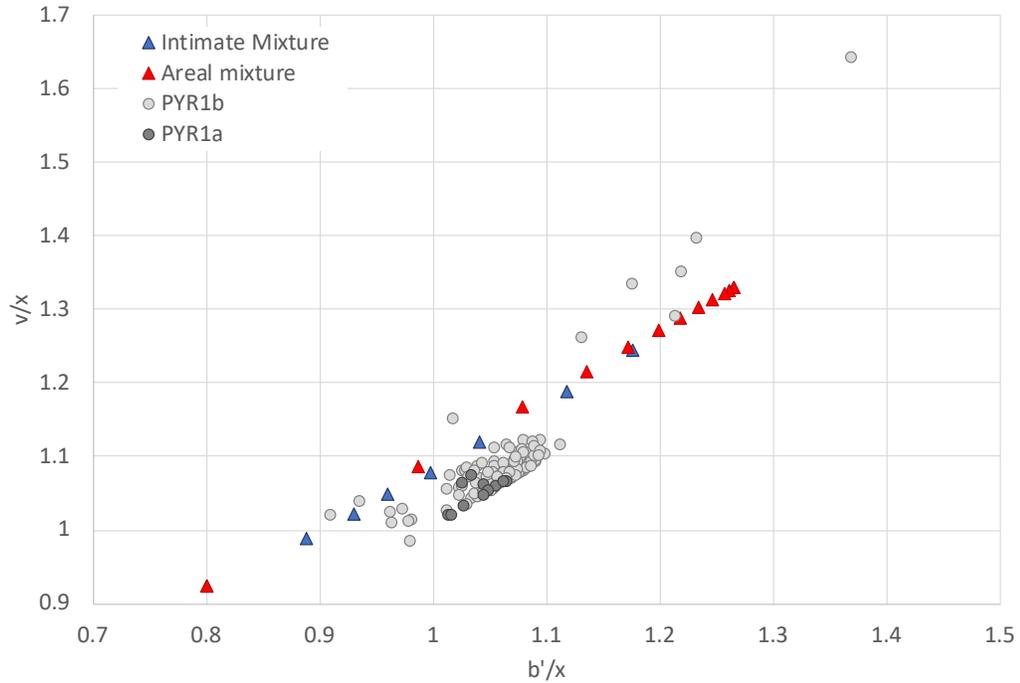

Next, we investigated whether exogenic boulders on Bennu have affinity to OCs, the dominant meteorite class in the inner solar system and near-Earth space. Figure 9A shows the average spectrum of the only boulder from the PYR1b subgroup with $W_{peak} > 10\%$ that was not examined with OVIRS data in DellaGiustina et al. (2020a), along with selected resampled laboratory spectra of H, L, and LL OC meteorites based on the best fit to this boulder. All these meteorite subtypes have higher v-band albedos (0.28—0.30) than this PYR1b boulder (0.09), despite having similar shapes of color spectra. One explanation for this could be that the laboratory spectra of meteorites were collected from powdered samples and hence have higher albedos compared to the bare rocks (slabs) on Bennu. Alternatively, the lower albedo could be the result of optical mixing with carbonaceous material due to dust coating on this OC-like boulder (as proposed for some eucrite-like boulders in DellaGiustina et al. (2020a)). For PYR1a, a selection of three impact melt OCs (from H, L, and LL) is the closest match to this group; the $W_{peak}$ is slightly higher but is better than any HED fit that we found. PYR1a boulders have a v-band albedo averaging 0.045, which is lower than the selected OC melts (0.09—0.18). Computing areal mixtures of the selected OC impact melts with the global average spectrum of Bennu (Figure 9B) provided a better fit to some PYR1a boulders (for example, with 10% L6 Mreira melt). Similarly, the best meteorite matches for the PYR1b boulders with $W_{peak} < 4\%$ are also H, L, and LL impact melts, but the best fits to some of the boulders are mixtures of some OCs with Bennu's average spectrum (Figure 9C). Using principal component analysis, Tatsumi et al. (submitted) also found that some of the exogenic boulders in their dataset may represent mixtures of OC meteorites with material having Bennu's average spectrum.

With the boulders selected in Figure 9, plus four additional boulders from the PYR1b sub-group that were not plotted in this figure for clarity, we can estimate the net amount of OC material potentially found on the surface of Bennu. Assuming that the selected boulders from PYR1a and PYR1b with $W_{peak} < 4\%$ are boulders made of pure OC material covered with low-albedo carbonaceous dust from Bennu, and that boulders have an ellipsoidal shape, we computed the total volume of OC material to be 5353 $m^3$. This value corresponds to the minimum amount of OC possibly present on the surface of Bennu based on interpretation of the four-color MapCam data. To get an estimate of the total amount of OC material present in the interior of Bennu, we assume that Bennu is uniformly mixed with debris, and therefore, what is observed on the surface is representative of the concentration of OC material in the interior. Assuming that the OC material resides in a shell around Bennu that is 3 to 5 m deep, the concentration of this material in the shell would be about 0.1–0.2%. The net volume of OC material in the interior of Bennu can then be estimated to be between 91,000 and 150,000 $m^3$.

*Figure 9.* MapCam color spectra of exogenic boulders from the different classes compared to the best fit resampled laboratory spectra of ordinary chondrite types (H, L and LL) or mixtures of OCs with Bennu average spectrum. (A) A boulder from the PYR1b subgroup with $W_{peak} > 10\%$ and H, L, and LL OC meteorites. (B) A subset of the PYR1a boulders compared to a mixture of the 10% L6 Mreira melt and 90% Bennu average color spectrum. (C) A subset of the PYR1b boulders with $W_{peak} < 4\%$ compared to mixtures of OC and Bennu average color spectrum.

A.

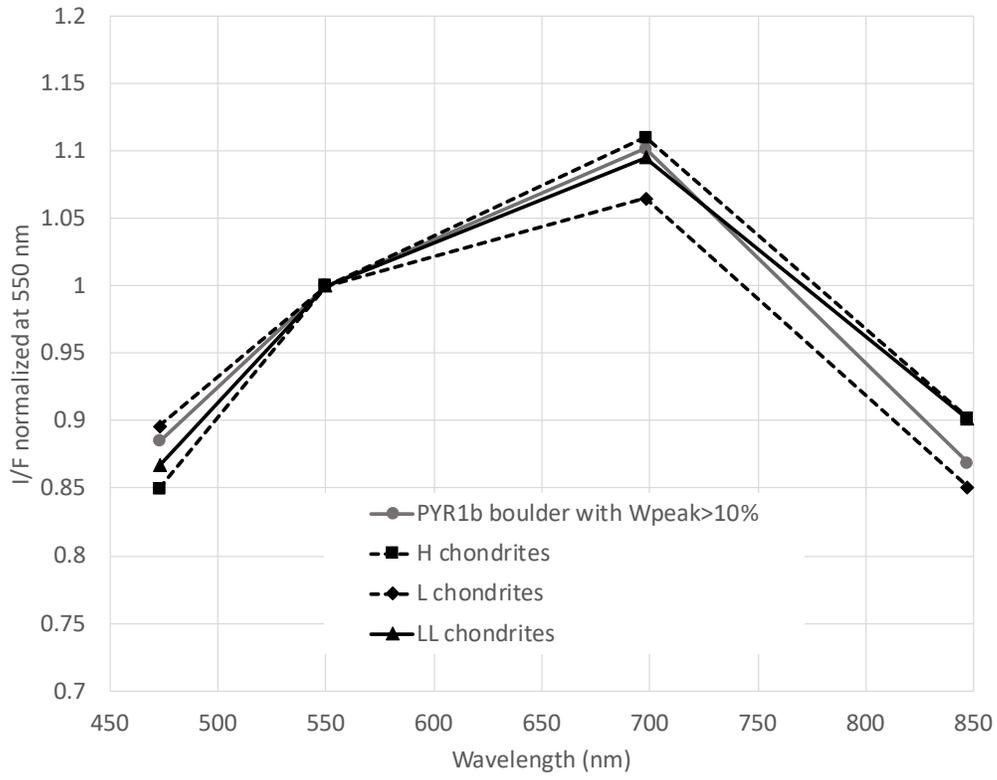

B.

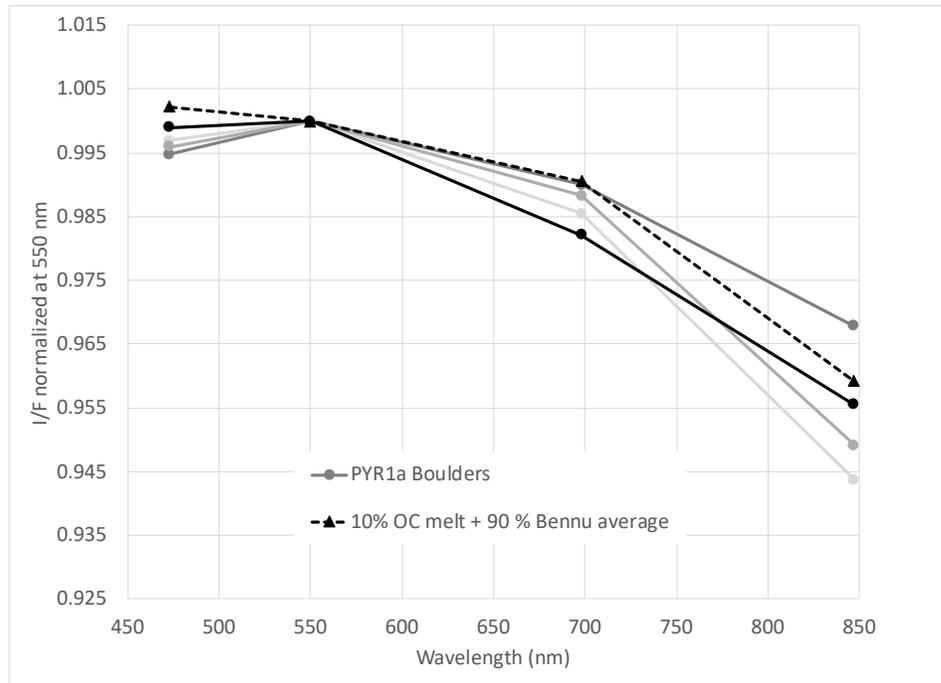

C.

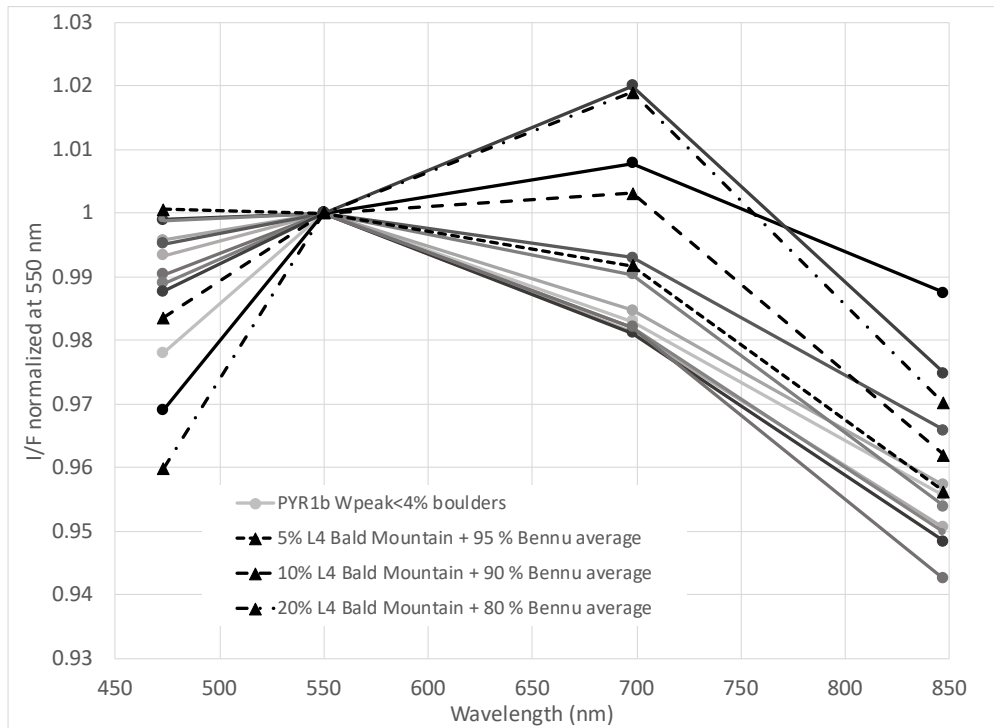

We further investigated the similarity of candidate exogenic boulders to ordinary chondrites by plotting color ratio scatterplots. Figures 10 and 11 show the ratios of b'/v and v/x color filters for the PYX1a and PYX1b boulders along with the three OC subtypes. We found that several outlying boulders in these color spaces match those of ordinary chondrites. 55% of PYR1a and 19% of the PYR1b boulders are distributed within the range of OCs in the b'/v vs. v/x plot (Figure 10). The rest of the PYR1a and PYR1b boulders are within the range of values for v/x observed for OCs, but their b'/v ratio is slightly higher than the OCs' range due to their flatter b'-v slope compared to OC spectra. The resemblance with OCs is also observed in Figure 11 (v/x vs. b'/x) where some PYR1a boulders plot close to the OCs, and ~10% of PYR1b boulders overlap with the OC clusters of points. PYR1a and PYR1b boulders also plot close to the OC trendline. Although four-band color data from MapCam is not as diagnostic of composition as hyperspectral data from the spectrometers onboard OSIRIS-REx (OVIRS and the OSIRIS-Rex Thermal Emission Spectrometer, OTES), it provides spectral information at a higher spatial resolution than either instrument. A follow-up study using the full spectral resolution and coverage of the OVIRS instrument should be undertaken to confirm our findings.

*Figure 10.* Scatterplot with the ratio of b'/v filters on the Y-axis and v/x filter on the X-axis for PYR1a and PYR1b boulders. Resampled laboratory spectra

of 57 OCs (H, L, and LL types) are also shown on this plot with different symbols. The color properties of some boulders are similar to those of OCs suggesting a compositional affinity.

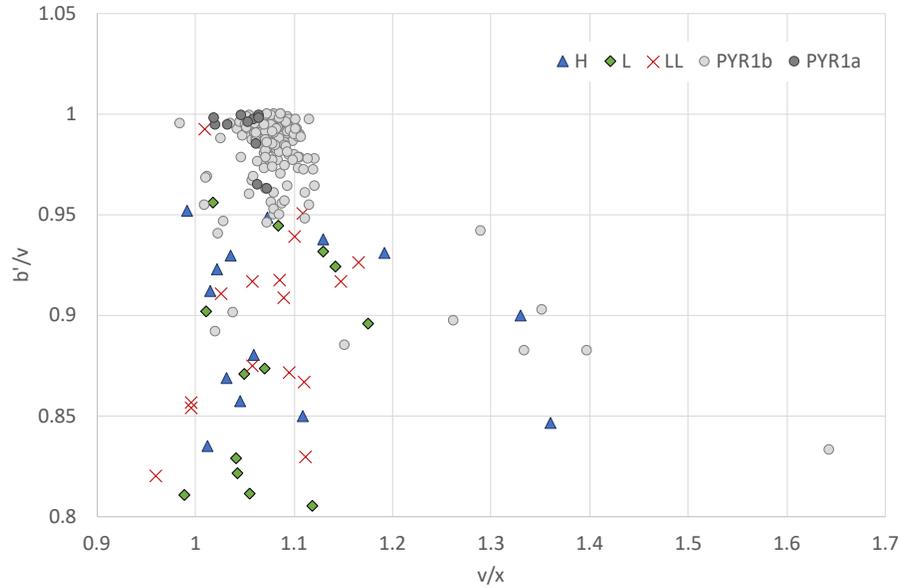

*Figure 11.* Scatterplot with ratio of v/x filters on the Y-axis and b'/x filter on the X-axis for PYR1a and PYR1b boulders. Resampled laboratory spectra of 57 OCs (H, L, and LL types) are also shown on this plot with different symbols. The color properties of some boulders are similar to those of OCs suggesting a compositional affinity.

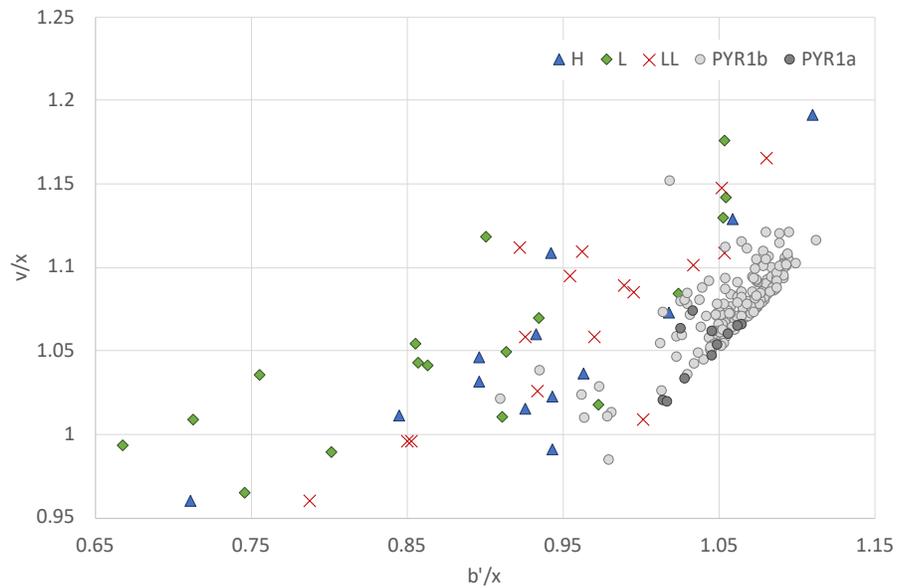

# 7. IMPLICATIONS

The possible resemblance of some pyroxene-rich boulders on Bennu to OC meteorites encouraged us to explore their source in the main asteroid belt. Using the concepts discussed in Bottke et al. (2015), we assumed that Bennu was created by the catastrophic disruption of a roughly 100- to 200-km-diameter parent body in the inner main belt (Campins et al. 2010; Walsh et al. 2013). The most likely candidates to have produced Bennu are the Eulalia and New Polana families, which have estimated ages of 830 [+370, −100] Myr and 1400 [+150, −150] Myr, respectively. If these assumptions are valid, there are several ways that Bennu could have been contaminated by ordinary chondrites:

1. Bennu's carbonaceous chondrite–like parent body may have been struck by OC asteroids (S-type) prior to its catastrophic disruption, and some of that exogenic material may have been incorporated when Bennu reaccumulated after the parent body's disruption.
2. Bennu's parent body may have been disrupted by an OC projectile, and some of that exogenic material may have mixed into the reaccumulated Bennu.
3. Bennu, or an immediate precursor, may have been hit by OC material during its transit from the main asteroid belt to its current orbit in near-Earth space (semimajor axis, eccentricity, and inclination of Bennu ($a, e, i$) are 1.126 au, 0.204, and 6.035º, respectively).

We address each scenario in sequence.

## 7.1. Adding S-type material to Bennu's parent body prior to its disruption

Bennu's parent body was probably hit by many different types of asteroids prior to its disruption, including S-types. If we assume that many of these impactors accreted to the parent body over billions of years, some portion of this material must have mixed into the fragments produced by the parent body's disruption. Moreover, if we assume that mixing during the parent body's disruption was 100% efficient, we can estimate the fraction of exogenic material that Bennu would have had when it or its immediate precursor was formed by this collision event.

This calculation requires knowledge of:

A. The orbit and size distribution of S-type material in the asteroid belt.
B. The collision probability of these objects with Bennu's parent body.
C. The size of Bennu's parent body and the interval that it was hit by the S-type population prior to disruption.

### 7.1.1. Component A

An important yet underappreciated issue about main belt collisions is that most asteroids can strike one another, even if the target of interest is located at the inner regions of the main belt (e.g., Farinella and Davis 1992; Bottke et al. 1994). Accordingly, when we consider S-type impacts on the parent body of Bennu, we need to evaluate the entire population of S-types in the main belt, since most of them have some chance of hitting the parent body. This is accounted for in the collisional probability calculation that we will discuss in Component B.

Our initial task is to identify S-type material in the main belt and determine its approximate orbit and size distribution. We estimated the former using the Wide-field Infrared Survey Explorer (WISE) diameter-limited catalog of main belt objects (e.g., Masiero et al. 2011). The WISE catalog provides albedos for numerous asteroids with diameter $D > 1$ km in the main belt, but it is incomplete. Partially this is because some small portions of the main belt went unobserved over WISE's mission lifetime, but also because WISE and other asteroid surveys cannot detect main belt bodies below a threshold size that approaches a few kilometers on average. For this reason, we limited our study to asteroids with $D \geq 5$ km, a population that is arguably complete within existing asteroid catalogs for the inner, central, and outer main belt regions.

Using this catalog, we assumed that S-type asteroids have albedos that broadly fit within a range between 0.1 and 0.3 (Masiero et al., 2011; Morbidelli et al., 2020). This range is slightly larger than that assumed by Mainzer et al. (2012) for S-type asteroids, which was 0.22 [+0.08, −0.08]. Here we employ the rationale used by Morbidelli et al. (2020), who pointed out that both Mainzer et al. (2012) and Wright et al. (2016) found discontinuities in the albedo distribution near the 0.1 and 0.3 boundaries. For the lower boundary, this arguably provides the means to discriminate between carbonaceous and non-carbonaceous bodies.

One problem with our method to identify S-types, however, is that there are numerous mid-albedo bodies beyond 2.95 au (J7:3 resonance). Many of these bodies are carbonaceous K-type asteroids within the Eos asteroid family. Given that albedo is our only discriminant of taxonomic type in the WISE database, these bodies can easily be confused with S-types. Fortunately, observational evidence suggests there are not a significant number of objects beyond 2.95 au with S-type albedos, color, spectra, or the like (e.g., DeMeo and Carry, 2014; J. Masiero, personal communication). For this reason, we limited our search for S-types within the WISE database to bodies with $a < 2.95$ au. Over this semimajor axis range, we found that between 11% and 16% of the main belt with $D \geq 5$ km to $D \geq 50$ km had an albedo within 0.1 and 0.3. More specifically, for $D \geq 5$ km and $D \geq 50$ km bodies, we found that 4580 out of 38,437 asteroids (11%) and 86 of our 535 asteroids (16%) passed these criteria.

Next, we estimated the size frequency distribution (SFD) for S-type bodies in the main belt. Although there are many ways to attack this problem, we chose the following approximation. Taking the main belt size frequency distribution preferred by Bottke et al. (2020), defined in their Figure 1 as SFD #6, we obtained the S-type size distribution by multiplying it by the factor 0.11 to 0.16. Essentially, we scaled the main belt SFD by these fractions.

### 7.1.2. Component B

Our next component is the collision probabilities of S-type material with Bennu's parent body. We computed these using the formalism presented in Bottke et al. (1994). It uses geometric arguments to estimate the probability that pairs of bodies with different semimajor axis, eccentricity, and inclination ($a, e, i$) values can hit each over all possible orbital orientations.

For simplicity, we assumed that Bennu's parent body had the same orbit as (495) Eulalia, the largest surviving remnant of the Eulalia family. The orbital parameters of (142) Polana, the largest remnant of the New Polana family, are nearly the same, so we expect little difference between its intrinsic collision probabilities $P_i$ and those for (495) Eulalia. The orbits of the S-type asteroids were drawn from the WISE database discussed above, with proper elements defined by Knezevic and Milani (2003).

Using Eulalia's proper ($a, e, i$) values of (2.487 au, 0.0989, 2.602°), we calculated a mean $P_i$ value for S-type WISE asteroids on Eulalia-crossing orbits of $6.63 \times 10^{-18}$ km$^{-2}$ yr$^{-1}$ for $D \geq 5$ km bodies to $7.2 \times 10^{-18}$ km$^{-2}$ yr$^{-1}$ for bodies of $D \geq 50$ km. These values can be multiplied by the S-type SFD computed in the previous paragraph.

### 7.1.3. Component C

Putting components A and B together, we can estimate how much S-type material hit the Eulalia parent body per year per square kilometer. The missing components are the size of the parent body, needed for the cross-sectional area of the parent body exposed to S-type projectiles, and how long the parent body was plausibly hit by this S-type population. Based on estimates from Bottke et al. (2015), we assume the Eulalia parent body was 100 km or 200 km in diameter.

Such a body can be disrupted by 18- to 49-km-diameter projectile, respectively, striking at 5 km/s (e.g., Bottke et al., 2020). By definition, the latter projectile was the largest one ever to strike Bennu's parent body. We assume the Eulalia parent body disrupted ~800 Myr ago (Bottke et al., 2015). The total time for which it could have been struck by S-types is approximately 4.56 Gyr — 0.8 Gyr = 3.76 Gyr. We also make the simplifying assumption that the S-type population has been in a steady state over this interval (e.g., see Bottke et al. 2015).

### 7.1.4. Model Results

Combining all the components, we created a Monte Carlo code that would use random deviates to add S-type projectiles to Bennu's parent body over timesteps of 10 Myr. We found that after 10,000 trials, the median amount of OC mass added to the Eulalia parent body was 0.055% and 0.037% of the volume of 100- and 200-km-diameter parent bodies, respectively. This means that if Bennu was a uniformly mixed byproduct of parent body and S-type projectiles, and that all impactor material stayed on the parent body, this is the fraction of contamination we would expect throughout its volume. The equivalent mass of OC material would be a sphere with a diameter of 36 to 40 m, which would correspond to a total volume of material between 24,200 and 33,600 m$^3$. The total amount of OC material in the interior of Bennu estimated based on the MapCam data is higher, between 91,000 and 150,000 m$^3$, corresponding to a concentration of 0.1–0.2%. These concentrations are modestly lower than predicted from the surface OC contamination of Bennu, but we consider them reasonably close given our model uncertainties.

If we assume that the differences between model and observation OC concentrations are meaningful, we can explain them several ways:
1. Bennu's parent body was not uniformly mixed, and Bennu likely originated from the upper regions of the parent body that had higher concentration of S-type debris.
2. Bennu's parent body was hit by projectiles larger than average.
3. The concentration of OC material on Bennu's surface is modestly higher than that in its interior.

### 7.2. Adding S-type material to Bennu's parent body during its disruption

For scenario #2 above, where Eulalia was disrupted by an OC parent body, we can reuse our Monte Carlo code from section 7.1. The difference is that now we will only accept model outcomes that produce a catastrophic disruption event produced by an S-type asteroid (i.e., one projectile has to hit Eulalia that has a diameter larger than the disruption threshold). Given that S-types only make up 11—16% of the main belt, this can be considered a modestly low probability event but one that cannot be ignored.

Rerunning our Monte Carlo code, and assuming Bennu is a uniform mixture of projectile and target, we find that the median contamination is between 3 and 5%, which is higher than our estimated concentration of 0.1-0.2% based on the MapCam data. We hesitate to rule out this scenario, however, because an unknown fraction of the projectile may have been lost in the disruption event. If enough disruptor mass was lost, this scenario could readily explain observations.

Note that this fractional value of exogenic material above is high enough to suggest that ample projectile material could be returned by sample collection via the OSIRIS-REx spacecraft. The odds favor that such exogenic samples would be fragments from some C-complex bodies, which dominate the main belt. We consider this the most likely scenario, with the existence of such material difficult to detect among the predominantly B-type rocks seen on Bennu's surface.

On the other hand, we find it unlikely that such a disruption event could involve eucrites (i.e., a eucritic projectile disrupted the Eulalia parent body). The largest eucrite-like body in the main belt today, other than Vesta, is (1459) Magnya, which has a diameter of roughly 17 km (Hardersen et al. 2004). All other V-type bodies are smaller than 10 km. Accordingly to our estimates above, a Magnya-size object could just barely disrupt the Eulalia parent body. However, the probability that this took place is exceedingly low. It would require that some kind of singular body in the main belt larger than Magnya happened to end its life by striking the Eulalia parent body. Using the collision probabilities for S-type asteroids above, we find that the average interval for a single object to strike Eulalia is $6 \times 10^{13}$ yr, many orders of magnitude longer than the age of the solar system.

### 7.3. Adding S-type material to Bennu or its immediate precursor

Finally, we consider scenario #3, the possibility that S-type material struck Bennu and that some of it remained on Bennu itself. This solution is difficult to quantify without a more sophisticated modeling effort because much depends on the evolutionary history of Bennu itself. Open questions are whether the precursor(s) of Bennu shed mass (and how much) as a result of rotational acceleration, whether Bennu is a $N$th-generation body originating from a series of larger precursors who experienced collisional disruption events prior to leaving the main belt, and how many impacts of all sizes did these precursors receive from S-type bodies.

For simplicity, we used the asteroid disruption law from Bottke et al. (2020) to estimate the size of the projectile needed to disrupt Bennu. If projectiles strike at 5 km/s, we estimate that a 7.5-m body is capable of disrupting the 495-m-diameter Bennu. The ratio of the volumes for these two bodies is roughly $3.5 \times 10^{-6}$. If Bennu did not disrupt, or only disrupted once, and the net mass added to Bennu by OC projectiles never exceeded the disruption mass, scenario #3 is less likely than #1 and #2 for producing S-type contamination.

A possible exception to this calculation would be that objects larger than the minimum disruptor size may have hit Bennu, and this could change the probabilities above. We will explore this possibility in future work.

## 8. SUMMARY

We investigated the color spectral properties and meteorite affinities of candidate exogenic boulders larger than 1 m on near-Earth asteroid Bennu using data from the OCAMS instrument on the OSIRIS-REx spacecraft. The key findings of our study are as follows:

- We confirm the presence of exogenic boulders with meteorite affinities to eucrites from the HED group derived from asteroid Vesta, as first published in DellaGiustina et al. (2020a). Mixtures of eucrites with carbonaceous chondrite material are also a possible match for some bright exogenic boulders. These results are consistent with the conclusions of Tatsumi et al.'s (submitted) study of boulders with diameter ~1 meter or smaller.

- Some of the exogenic boulders also have spectral properties similar to those of ordinary chondrite meteorites, although the laboratory spectra of these meteorites show a higher albedo than those measured on Bennu. We attribute this difference to particle size effects, but optical mixing with carbonaceous material due to dust coating could also produce it. Mixtures of OCs with material having Bennu's average spectrum were also suggested for boulders smaller than 1 m by Tatsumi et al. (submitted). OC impact melts, which are relatively rare among the OCs, are a possible match for the darker exogenic boulders. Based on this study, the approximate net volume of potential OC-like material we identified on Bennu is 5353 m$^3$. This is more than the amount of material (~70 m$^3$) with an HED-like affinity estimated from the six boulders found by DellaGiustina et al. (2020a).

- We explored several different ways for Bennu to be contaminated by OCs. The most plausible mechanism is a collision of S-type asteroids with prospective parent bodies for Bennu. Out of 10,000 Monte Carlo trials, models indicate the median amount of OC mass added to a 100- and 200-km-diameter Bennu parent body corresponds to 0.055% and 0.037%, respectively, of their volumes. This is slightly lower than the amount of OC-like material we have estimated on Bennu based on the MapCam data (0.1–0.2%). We proposed several mechanisms to account for this difference. An alternative explanation for observations of OC material on Bennu would be the disruption of Bennu's parent body by an S-type asteroid and subsequent mixing.

- A more thorough investigation involving the analysis of data with higher spectral resolution and wider wavelength range from the OVIRS visible and near-infrared spectrometer, coupled with more sophisticated dynamical modeling, is necessary to confirm our findings.

## 9. ACKNOWLEDGMENTS

This material is based upon work supported by NASA under contract NNM10AA11C issued through the New Frontiers Program. This research has made use of the USGS Integrated Software for Imagers and Spectrometers (ISIS). We thank the entire OSIRIS-REx Team for making the encounter with Bennu possible. OCAMS MapCam and PolyCam data from the Detailed Survey—Baseball Diamond phase of the OSIRIS-REx mission are available via the Planetary Data System (Rizk et al., 2019). Shape models of Bennu, including v28, are available in the Small Body Mapping Tool at http://sbmt.jhuapl.edu/. This research utilizes spectra acquired with the NASA RELAB facility at Brown University (http://www.planetary.brown.edu/relab/).